\documentclass[runningheads]{llncs}
\usepackage[T1]{fontenc}
\usepackage{graphicx}
\usepackage{tabulary}
\usepackage{booktabs} 
\usepackage{graphics}
\usepackage{graphicx}
\usepackage{blindtext}
\usepackage{siunitx}
\usepackage{multirow}
\usepackage{url}
\usepackage{xcolor}
\usepackage{courier}
\usepackage{subcaption}
\usepackage{cleveref}
\usepackage{capt-of}
\usepackage[justification=centering]{caption}
\usepackage{makecell}
\usepackage{tabularx}
\usepackage[export]{adjustbox}

\newcommand\Tstrut{\rule{0pt}{2.6ex}}         
\newcommand\Bstrut{\rule[-0.9ex]{0pt}{0pt}}   

\usepackage[ruled]{algorithm2e} 

\SetAlFnt{\small}
\SetAlCapFnt{\small}
\SetAlCapNameFnt{\small}
\SetAlCapHSkip{0pt}
\IncMargin{-\parindent}

\newcommand{\eg}{\textit{e.g.},\xspace}
\newcommand{\ie}{\textit{i.e.},\xspace}
\newcommand{\etal}{\textit{et al.}\xspace}

\newcommand\showcol{0}
\newcommand{\hl}[1]{%
    \ifnum 1=\showcol \relax
        {\color{red}#1}\else #1%
    \fi
}

\begin{document}
\title{MarcoPolo: A Zero-Permission Attack for Location Type Inference from the Magnetic Field using Mobile Devices}
\titlerunning{MarcoPolo: A Zero-Permission Attack for Location Type Inference}
\author{Beatrice Perez\inst{1} \and
Abhinav Mehrotra\inst{2} \and
Mirco Musolesi\inst{3,4}}
\authorrunning{Perez, Mehrotra, and Musolesi}
%
\institute{University of Massachusetts, Boston, MA, United States of America  \\
\email{beatrice.perez@umb.edu}
\and
Samsung AI Center, Cambridge, United Kingdom \\
\email{a.mehrotra1@samsung.com} \footnote{This work was carried out by Beatrice Perez and Abhinav Mehrotra when they were affiliated with University College London.}\\
\and
University College London, London, United Kingdom\\
\email{m.musolesi@ucl.ac.uk}
\and
University of Bologna, Bologna, Italy
}

\maketitle              
\begin{abstract}

Location information extracted from mobile devices has been largely exploited to reveal our routines, significant places, and interests, just to name a few. Given the sensitivity of the information it reveals, location access is protected by mobile operating systems and users have control over which applications can access it. We argue that applications can still infer the coarse-grain location information by using alternative sensors that are available in off-the-shelf mobile devices that do not require any permissions from the users. 

In this paper we present a zero-permission attack based on the use of the in-built magnetometer, considering a variety of methods for identifying location-types from their magnetic signature.
We implement the proposed approach by using four different techniques for time-series classification. In order to evaluate the approach, we conduct an \textit{in-the-wild} study to collect a dataset of nearly 70 hours of magnetometer readings with six different phones at 66 locations, each accompanied by a label that classifies it as belonging to one of six selected categories. Finally, using this dataset, we quantify the performance of all models based on two evaluation criteria: (i) leave-a-place-out (using the test data collected from an unknown place), and (ii) leave-a-device-out (using the test data collected from an unknown device) showing that we are able to achieve 40.5\% and 39.5\% accuracy in classifying the location-type for each evaluation criteria respectively against a random baseline of approximately 16.7\% for both of them. 

\end{abstract}

\section{Introduction}
Mobile phones are equipped with a variety of sensor used by applications to obtain a user's contextual information (\eg  location, humidity, acceleration, and network connectivity) to support a variety of applications and services~\cite{internetcomputing}.
However, the sensors embedded in smartphones have also unintentionally become the source of information leaks that might adversely impact the privacy of their owners.
Indeed, information extracted through sensors can be used to identify users or devices~\cite{dey2014accelprint,perez2019magID,zhang2019sensorid,kumar2023device,mohamed2023istelan,yapar2023real}, and infer their behavioral patterns, interests, personal preferences~\cite{noulas2009inferring}, and even their health condition~\cite{zhou2013identity}. The fact that information can automatically be extracted by means of passive sensors is generally perceived negatively by users~\cite{brush2010exploring,cvrcek2006study,shih2015privacy,song2010limits}.

In order to mitigate the privacy risks, mobile operating systems have adopted a permission-based paradigm where each application must request access to any of the protected resources on the phone, including camera, microphone, location, and contact list.
Each permission is flagged as sensitive depending on the invasive nature of the resource and the importance of the data it might reveal; therefore, the permission to answer phone calls is more restricted as compared to the permission \hl{that allows for access to the camera of the device}. Users can choose to disable system-wide access to a certain type of information, or control it at the application level where access to some information can be revoked. 

Location is one of the information types that is protected by such permission systems. Location data has been shown to be sensitive for users as it can be used to track and profile individuals (\eg for advertising purposes). In particular, it can be used to infer a user's identity~\cite{deMontjoye2013unique}; 
it can reveal a user's significant locations and points-of-interests (\eg home location, work location, morning coffee shop, events attended) along with their transportation routine and use them to predict trajectory and future locations~\cite{kautz2004learning}; and finally, it can be used to detect the general behavior of a single user and group users based on the similarity across their interests~\cite{li2008mining}.

Restrictions placed on location data have led to the development of alternative methods that aim to derive location from sensors not protected under permission systems. Currently, Android and iOS application can access the gyroscope, accelerometer, and magnetometer sensors without requiring the user permission. These methods generally combine previously acquired environmental information (\eg produced by surveying a location) with real-time sensor readings to ascertain the user's current position~\cite{Wang2010survey}. In particular, Wang~\etal combine the accelerometer and gyroscope in order to localize users 
~\cite{wang12wardrive} and Chen~\etal combine WiFi signal strength with FM radio signals to determine a users' location~\cite{chen2012fmloc}. 

\begin{figure*}
    \centering
    \includegraphics[width=1.0\textwidth]{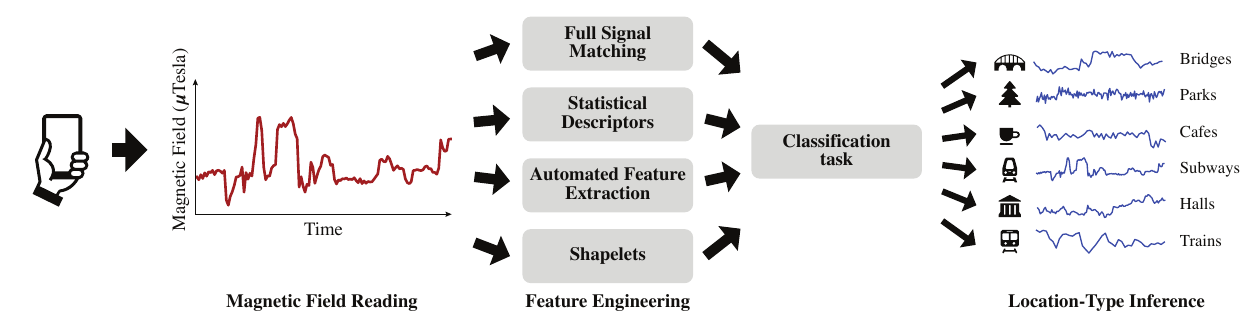}
    \caption{Overview of the zero-permission attack leveraging the magnetometer of mobile devices to capture magnetic field readings.}
    \label{fig:overview_attack}
\end{figure*}

In this paper, we show how one can leverage one of these sensors, which is not subjected to any permission, in order to infer the type of the location where the user is currently situated.
We are not the first to propose such a solution. However, where previous studies focus on mapping locations to pinpoint position, we present a method that reduces the need for surveying the target locations (\ie locations where localization will take place).
We also require that the identification process relies exclusively on data that can be accessed without permissions, thus devising a \textit{zero-permission localization attack}.
Our approach exploits the readings captured by the magnetometer, present in most smartphones today, to infer a type of location from their magnetic signature. 
We present four different methods for time-series analysis of magnetic readings: (i) full signal matching, (ii) statistical descriptors, (iii) automated feature extraction, and (iv) shapelets analysis~\cite{ye2009time}.
We compare the two traditional methods of statistical features and full signal pattern matching with two novel methods: convolutional neural networks through One Shot Learning and shapelets, that is a short time-series capturing repeated patterns observed in magnetic field readings.
The use of different techniques for analyzing time-series data enables us to establish baselines and compare their performance when applied to this problem. 
An overview of the attack is presented in Figure~\ref{fig:overview_attack}.
In order to evaluate our approach we collect a labeled dataset, used as ground truth, by conducting an in-the-wild measurement study and sampling the magnetic field at different locations over a metropolitan area. We collect readings from off-the-shelf smartphones at ten different types of locations: long-distance train stations, urban train stations (\ie subways), parks, bridges, coffee shops, halls, laundromats, bus stops, parking lots, and gyms. The places were chosen as alternatives for three groups: indoor environments, outdoor environments, and environments that contain clearly distinguishable events (\eg trains passing by). 
For each of the 10 location-types, we collect 10 minutes of magnetic field readings sampled at 1 Hz from five different phones at 11 locations. The entire dataset therefore consists of a total of 91 hours of readings. We evaluate the prediction performance of our four methods by first identifying a location-type from the magnetic readings of an unknown location and second, by identifying a location-type from the magnetic readings of an unknown device. 
We summarize our main contributions as follows:
\begin{itemize}
    \item We collected over 91 hours of labeled magnetic field readings from various locations across a major metropolitan area. We show that location-types can be inferred from magnetic field readings available without any system permissions across a wide range of smartphones.
    \item We compare the performance of four methods for time-series classification in identifying location-types. We perform an in-depth analysis of the choice of the values of the parameters for each methods, providing a methodology for their selection.
    \item Through our proof-of-concept implementation and in-the-wild study across different locations, we demonstrate the feasibility of the location attack.
\end{itemize}



\section{Overview}
\label{overview}

In this section we introduce the key concepts and present the main motivations for our work.

\subsection{Key Concepts}
\label{definitions}
\noindent In writing this work, we take ownership of some words and create new concepts. We define them as follows:

\smallskip\noindent\textbf{Location-Type.}
As defined in~\cite{thrun2005probabilistic}, localization is the problem of ascertaining the position of an agent relative to a map. In this work, we address the problem of localization from a novel perspective, where our primary interest is to understand the environmental similarities between places.
Location-type is the name we assign to the top level of a two-tier hierarchy (the lower level being the distinct places). Locations are grouped based on common environmental characteristics including, building structure and materials, human movement patterns, and events (defined later in this section).

\smallskip\noindent\textbf{Magnetic Field.}
The magnetic field is the combination of geomagnetic and electromagnetic phenomena. Geomagnetic fields describe the naturally occurring (magnetic) field emitted by the planet's core and the local variations caused by ferromagnetic materials such as iron, cobalt, or nickel~\cite{gozick11magnetic,haverinen09global}. As an example, the steel structure of a building and the movement of metal objects (\eg a car or a train) will distort the planet's field. 

\smallskip\noindent\textbf{Time-Series.}
A time-series is an ordered set of values (\ie the magnetic field) sampled at a fixed interval that, together, form a single observation or reading~\cite{fu2011review}. The length of the time series is equal to the number of values available in that observation. 
Subsets or samples of a time-series with more than one element (which will be associated to shapelets later) are time-series observations in their own right, with the distinction that the length of the subset must be no greater than the length of the originating observation(s). 


\smallskip\noindent\textbf{Events.}
We define events as the dynamic extension of landmarks for time-series data.
While a landmark represents a set of structural characteristics of an environment (\eg the corner of a building, the presences of a fountain, the electrical wiring of a room), an event represents a transient occurrence (\eg the movement of a train or car, riding on an escalator, or walking by a person). Unlike landmarks, which are unique to a place, events can be identified in places described by a location-type. 
In our analysis, we make the assumption that locations that belong to the same type are characterized by similar events, which will be associated to specific \textit{shapes} in the time-series of the magnetic field.

\subsection{Motivation}

Information about a user's physical location provides private insights about the users themselves. Indeed, applications installed on the users' smartphones can access location data provided by location services using the GPS chip and the cellular network through permission-based controls offered by the operating system. It is left to the user's discretion to grant location access to any application that requests it~\cite{ApplePermission,androidPermissionLevel,WindowsPermission}. 

On the other hand, side-channel, zero-permission attacks have been exploited to infer users' location using alternate approaches without requesting access to the corresponding GPS and cellular permissions. They rely on constructing, for each target location, a map of the environmental characteristics measured by the sensors embedded in the smartphone. These sensors, whose access does not require any specific permissions, include primarily the magnetometer, the accelerometer and the gyroscope~\cite{han2012accomplice,michalevsky2015powerspy,narain2016inferring,nawaz2014mining,zhang2013senstrack}. In practice, given a target location (\eg an area or a building), a map can be built by recording sets of sensor measurements (\eg acceleration from the accelerometer and/or magnetic field from the magnetometer) that are associated to precise coordinates within the location being surveyed. 


As opposed to the GPS module, the magnetometer is a low-power sensor and accessing readings from a magnetometer does not require any notification to users (or permissions) in either Android or iOS.  Moreover, once the information is processed it can be used to protect the privacy of users: while the GPS provides detailed location information (accurate to centimeters of the true position), a generic localization method based on magnetic field could potentially be used to verify that a user is within the premises of a certain type of place or in a specific environment without disclosing exact information. Both marketing and verification applications could obtain the information they need without being invasive with regards to user location.
\section{Related Work}
\label{related}

Our work is related to the general problem of localization. According to the taxonomy presented by Thrun~\etal, we address a global localization problem in a dynamic environment where we passively monitor users~\cite{thrun2005probabilistic}. 
In this context, localization of an autonomous entity, referred in the literature as ``agent", can only be carried out through sensor readings, either by matching current readings against an up-to-date map or through geometric calculations that determine the unknown position of the agent against a known marker. In the following, we discuss different approaches and contextualize this work with respect to the literature. 


\subsection{Model-Based Localization}
Model-based techniques include methods where the movement of the user is calculated from a sensor, including GPS, as well as cellular and WiFi signals. They are by far the most common and wide spread techniques used for localization. 

Mobile operating systems give developers the option to localize users based exclusively on cellular radio signal strength. 
The position of the user is inferred by extracting features from a signal and comparing them to an established ground truth. As an example, network signal strength is one of the most commonly used features and the actual position of the user is computed from the triangulation between multiple cellular base stations and the signal received from each station by the device~\cite{vo16outdoorSurvey}.

These methods can be applied to indoor environments as well. Localization from WiFi access points using received signal strength has been found to be accurate enough to track users inside buildings with the capability of distinguishing between adjacent rooms~\cite{chang10wifi,chintalapudi2010nopain,yang09rss}.
The method only requires WiFi signal strength and the layouts of the building so as to localize a user. 
The primary limitation of these methods is that the signal strength calculation is typically carried out on the device and continuous tracking drains the resources of a phone~\cite{chintalapudi2010nopain}.

Model-based techniques are used as a means of error correction for dead-reckoning schemes. Urban dead-reckoning is the process by which a user's location is tracked by measuring the side-effects of motion, usually through the use of the compass and the gyroscope. It has been shown that urban dead reckoning accumulates errors of up to 100 meters in 6 minutes of collected data~\cite{wang12wardrive}. Indoor localization methods often combine dead reckoning with model-based localization to maintain acceptable tracking accuracy~\cite{woodman08pedestrian,wu13will}. 

The error correction mechanism is obtained from a variety of sources. For example, Haverinen and Kemppainen use GPS readings to mark the entrance of the building and again collect active GPS readings from open areas or windows to validate and correct their location~\cite{haverinen09global}. Wang~\etal collect magnetic field readings along with gyration and acceleration information from different users to find landmarks and correct the error for individual users~\cite{wang12wardrive}.

\subsection{Fingerprint-Based Localization}
Fingerprint-based techniques include all methods where a site (or any other geographical location) is surveyed in order to build a map which is then used to pinpoint the location of a user. Fingerprint-based techniques are costly with respect to collection and maintenance of the maps. Active areas of research in this field are mainly centered on reducing the overall financial cost necessary for building maps primarily through leveraging crowd-sourcing techniques where data is passively collected by a large population~\cite{rai2012zee}. In the following, we detail the approaches specific to the type of environment, whether it is indoor or outdoor. This type of techniques have been used successfully to fingerprint devices themselves (see, for example, \cite{perez2019magID,zhang2019sensorid}).

\smallskip\noindent\textbf{Indoor Environments.}
Localization in an indoor environment is linked to a map with high level of detail. Even for small spaces this presents a challenge in terms of the volume of data that must be available for the task. In fact, it is important to consider that sensor values might vary with altitude as well as with horizontal displacement. Furthermore, selecting a sensor that provides consistent differences in enclosed environments increases the complexity of the task. 
Potential solutions to these problems are manual collection of ambient readings or strategic placement of location beacons (and sensors) throughout the area of study under consideration~\cite{chung11indoor,gozick11magnetic,wang12wardrive,wu13will}.
Chung \etal use the magnetic field to determine the position of a user in a corridor. They manually collect magnetic field readings every 60 cm and find that localization is accurate to 1.64 m for 90\% of the test observations~\cite{chung11indoor}. They also test their system in elevators and in the atrium of a building and find that there are measurable differences across these locations. 
Following the work by Chung \etal, Haverinen and Kemppainen test whether the difference in magnetic field readings can be used to locate an agent across a larger area (in their work, they test four buildings). They find that these readings have low variability over time while being spatially distinct and build a localization system based on them~\cite{haverinen09global}. 
Finally, Wang~\etal show that magnetic field readings can be used to discriminate between activities such as standing or walking. In particular, they use changes in the magnetic field to identify when a user is moving in an escalator~\cite{wang12wardrive}.

\smallskip\noindent\textbf{Outdoor Environments.}
Outdoor localization approaches are classified based on the technique used to establish the map. 
One common fingerprinting method from outdoor environments is matching the visual cues obtained from the camera of the phone to a map of geo-tagged images~\cite{vo16outdoorSurvey}.
Narain~\etal use the combination of the motion and position sensors available in smartphones (primarily the accelerometer and gyroscope) to infer the trajectory of a moving car~\cite{narain2016inferring}. In particular, they propose a method to reconstruct the route taken by the car based on the physical characteristics of the roads (\ie speed, bumps, stop signs, and curvature), the junctions between roads, and the angle of each turn was recorded. With this information, they are able to match the estimated candidate trajectory to the map of the (known) city using OpenStreetMap and estimate the actual route taken by the user. Similarly to our work, the authors propose a zero-permission attack to determine location information about users from motion sensors. However, we focus on a different localization task that aims to identify the places that users have visited solely from the magnetometer sensor instead of the trajectories they have taken.
Moreover, we employ an event-based approach that identifies events and matches events to locations thereby reducing the need for up-to-date maps. 
Finally, \cite{block_magnetometer}, the authors present a system where a location ID is transmitted via low power magnetic coil and received by permissionless apps. This system is different from ours, since MarcoPolo does not rely on any external device for identification.

\subsection{Contribution to the State of the Art}

The main challenge faced by outdoor localization techniques comes from the amount of data required for accurate comparison as well the ability to search this data efficiently in order to find the user's current location. On the other hand, the main challenge of indoor localization techniques is the design of fingerprinting methods that are costly in both manpower and time because they require fine spatio-temporal granularity~\cite{wu13will}. Moreover, the surveyed signal used to construct the map may change over long periods of time. Consequently, maps have to be repeatedly updated. Similarly, the calibration process on the device that is necessary to accurately measure the signal and the calculation of distances between devices and (signal) sources requires complex computations that might adversely impact the performance of smartphones~\cite{wang12wardrive}. 

To the best of our knowledge, our work is the first that explicitly maps location-types in terms of their characteristics, such as the structural similarity between buildings, the isolation of repeated events, the size and distribution of people and crowds, and any similarity across the tasks undertaken at each location. We combine these characteristics to generate a \textit{magnetic signature} for each location-type which we exploit to identify new locations in the city.
The signatures for each location-type are composed of a set of shapelets as defined in~\cite{grabocka2014learning}~\cite{hills2014classification}~\cite{mueen2011logical} ~\cite{ye2009time}~\cite{zakaria2012clustering}. The classification of a new observation into a location-type is a function of the likelihood of a shapelet being present in that observation.
To summarize, we propose a novel, low-cost, and passive methodology for location inference. In particular, we present supervised classification algorithms that take as input the time-series (observations) for all classes and predict the probability and label (\ie location-type) of a new observation.
\section{Methodology}
\label{method}

In this section we detail the different approaches that we use in our methodology to predict location-types from magnetic field readings. In particular, our methodology addresses two problems: (1) classification from time-series; and (2) deriving location-types from magnetic field readings.

\subsection{Techniques for Feature Extraction}
\label{sec:feature_models}

Measurements over time provide a more complete view of reality, they allow us to uncover relationships between consecutive measurements. We propose that collecting and processing longitudinal magnetic field readings will prove to be useful in the task of localization. However, time-series classification, clustering, and prediction are identified in the literature as hard problems. Challenges in time-series analysis include: depending on the length of the time series, resource consumption (\ie computational complexity) of the method; definition of a similarity metric in data that is noisy and prone to outlying values and shifts; and finally, high-dimensionality of observations. An in-depth discussion of these challenges can be found in~\cite{zakaria2012clustering,fu2011review,aghabozorgi2015clustering,bagnall2017great}.
In this work we compare four methods for time-series analysis: (i) full signal matching; (ii) statistical descriptors; (iii) automated feature extraction through the use of neural networks; and (iv) classification through the extraction of shapelets, in both time and frequency domains.
Details of these methods are discussed in the remainder of this section.

\subsubsection{Full Signal Matching.}
Pattern matching is a method for time-series classification where each test sample is matched against all labelled training samples. The test sample is then assigned to the label of the closest training sample (\ie with the shortest distance to it). Consequently, this approach requires a pairwise distance calculation from every signal in the testing set to every signal in the training set, which might become intractable as the number of signals grows. 
In this analysis, we used four different distance measures for performing classification task: Dynamic Time Warping (DTW), Euclidean Distance, Cosine Distance, and Bhattacharyya Distance~\cite{kailath1967divergence}. We selected these distance measures as they extract different information from the corresponding input signals: DTW involves both signals and tries to find the best fit between them in a cross-correlation computation, the Euclidean Distance measures the magnitude of the separation between the sampled signals, the Cosine Distance measures the angular separation between the two vectors being compared, and finally, the Bhattacharyya Distance captures the divergence between the probability distributions of each of the signals. One of the key limitations of this approach is that it is sensitive to small variations between the two signals, including differences in the mean for matching signals, which could negatively impact the classification performance. In contrast, we aim at identifying similarities between observations collected in noisy environments. Therefore, in the context of our application, we are expecting this method to have poor prediction accuracy. Nevertheless, we use it as the baseline metric for our performance evaluation.

\subsubsection{Statistical Descriptors.}
The second approach we consider consists in the extraction of representative statistical descriptors (i.e., features) from the time-series. The extracted features are used to represent key characteristics of the data. It is worth noting that the classification performance using these features depends on the degree to which the sampled data represents the population of interest. Good statistical models require a sufficient sample size.
We selected eight commonly used statistical features: the median, the amplitude, the energy of the signal, the magnitude and frequency of the natural frequency, the spectral centroid, and the magnitude and frequency of maximum power. Note that we sub-sample the input data and compute these features in both the time and frequency domain. It is worth noting that other alternative features might be extracted. 
Also for ensuring the replicability of the experiments, we used the open source versions of three state-of-the-art classification algorithms provided by the scikit-learn library~\cite{pedregosa2011scikit} to construct our prediction models: $k$-Nearest Neighbors ($k$NN)~\cite{cover1967knn}, Random Forests (RF)~\cite{breiman2001randomForest}, Extreme Gradient Boost (XGB)~\cite{chen2016xgb}. $k$NN is one of the most popular and effective unsupervised learning algorithms, whereas RF and XGB are widely used for their interpretability and performance, respectively.

\subsubsection{Automated Feature Extraction.}
Algorithms based on neural networks remove the burden of manual extraction of features in order to train prediction models~\cite{goodfellow2016deep}. 
In particular, in this study, we use Siamese Networks \cite{koch2015siamese}~---~a specific type of neural network architecture that aims to learn to differentiate between two inputs rather than classifying the inputs into given classes. This network architecture is comprised of two identical neural networks each taking one of the two inputs and the last layers of the two networks are then fed to a contrastive loss function to compute the similarity between the inputs~\cite{hadsell2006dimensionality}.  
Two neural networks are identical if they have the same configuration in terms of parameters and weights.
Since the weights across the two networks are shared, there are fewer parameters to train, which in turn means that less amount of training data is required. This also reduces the chance of overfitting. 
Since our input is in the form of a time series, we use 1-D convolutional layers to extract patterns from the data, which are passed from the feed-forward layers to obtain the results for the output layer that is used for computing similarity. More specifically, our network consists of $C$ convolutional layers (CNN layers) and a feed-forward layer that maps the features (extracted by the CNN layers) to the output layer with 100 nodes (\ie the number of final features to be extracted). We considered values of $C \in [2,3,4]$. We do not consider higher values of $C$ given the size of the training set under consideration. 

\subsubsection{Shapelets (in Time Domain).}
\label{method_shapelet}
\begin{figure}[t]
    \centering
    \includegraphics[width=0.9\columnwidth]{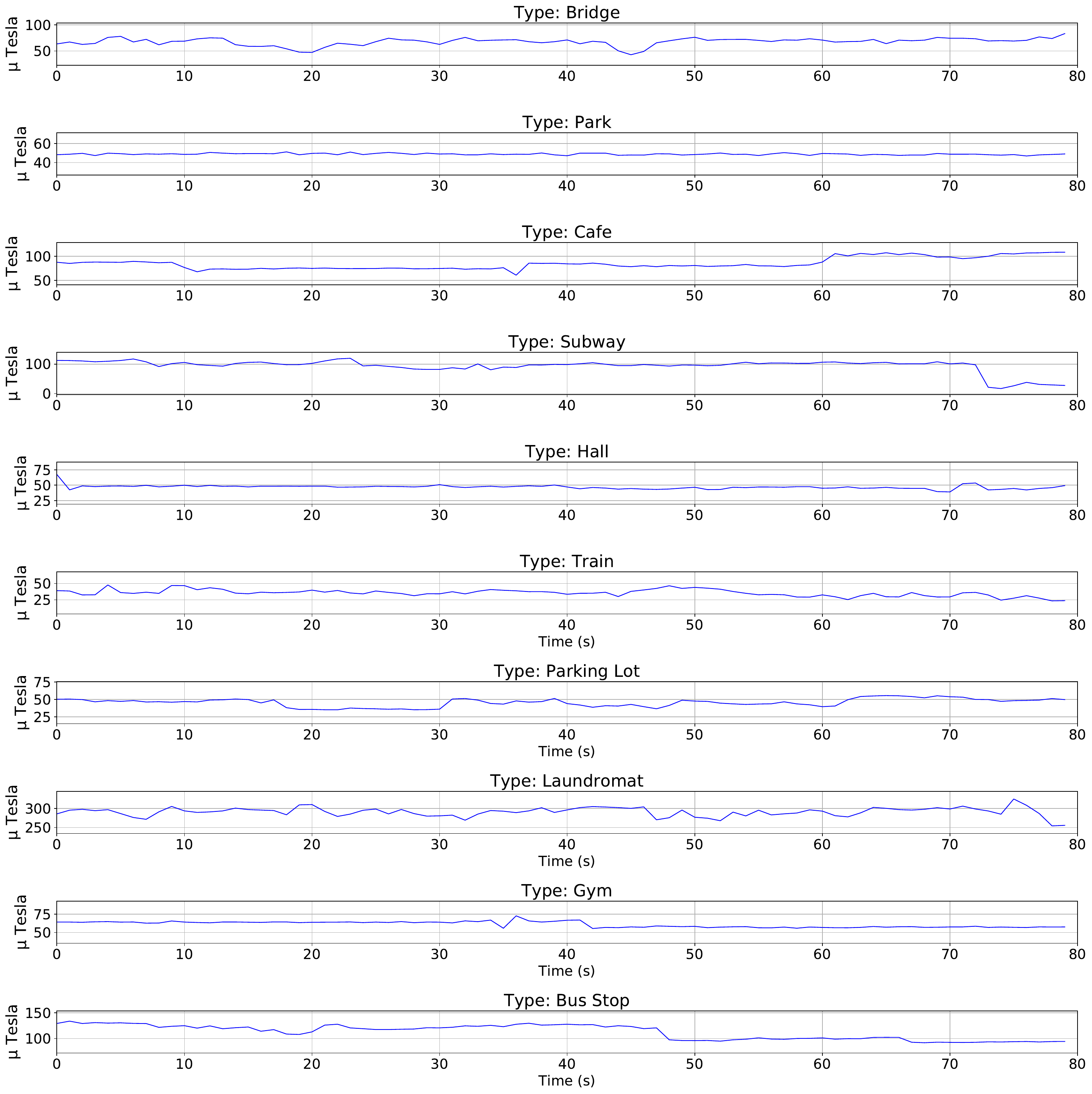}
    \caption{Visual representation of a \textit{shapelet}. In the figure, each sequence is the longest common time-series present in at least 90\% of the observations for each location-type. The length of the shapelet can vary per class as it is possible to observe in the plot. Some might be shorter than others.}
    \label{plot:shapeletExample}
\end{figure}

Shapelets are small local patterns in a time-series that are highly predictive of a class~\cite{mueen2011logical}. 
In the context of our work, each shapelet corresponds to an \textit{event} that is found in at least 90\% of the observations belonging to a certain \textit{location-type} (with 11 distinct locations and 5 devices we require an event to be present in 50 observations before a candidate is considered a shapelet).
Our assumption is that for each class there exists at least one shapelet contained in all observations of that class, and we consider shapelets of different sizes where longer shapelets are more valuable in terms of class separation.
%
One benefit of this technique is that the analysis of candidate shapelets is independent for each class. This means that the length of a shapelet is also a feature we are considering in the input and that the maximum length is determined on a per class basis (\ie one way to distinguish between classes that may have similar events is the duration of that event). As an example, the movement of a long-distance train and a subway are similar events but long-distance trains are longer which might correspond to longer shapelets.
The methodological contribution of this work incorporates shapelets into probabilistic classification algorithms.
Our method takes labeled time series observations and returns the location-type of an unlabeled observation. The algorithm is divided in three steps: first, it extracts shapelets from the training observations (to build a shapelet dictionary); then, uses the training data to describe each class in terms of the discovered shapelets; and finally, it classifies test observations into the known classes. 

\subsubsection{Shapelets (in Frequency Domain).}

It is common practice in signal processing to study signals in the frequency domain~\cite{brook1988signal}.
Following the same procedure described in Section ~\ref{method_shapelet}, we generate a bag of shapelets (with all the possible sub-signals from the training set) and transform each one to the frequency spectra before computing the correlation between all signals. 
Doing the piecewise transformation before extracting the shapelets might result in better classification if the sources of the signal are monotonic. 
We integrate the frequency analysis method, known as the Generalized Correlation Coefficient (GCC)~\cite{rabiner1978digital} with the shapelet-based classifier and compare the accuracy of both methods. 
We follow the same clustering procedure and apply hierarchical clustering to the distance matrix generated using GCC. The shapelet dictionary is extracted from the inspection of each cluster. During classification, we use GCC to compute the correspondence between each of the elements in the shapelet dictionary and the signal used as input. This process is repeated (separately) for the observations in the test set and evaluated using the same algorithms.

\subsection{Criteria for Assessing Prediction Models}
\label{sec:evaluation_criteria}


\subsubsection{All Places, All Devices.}
This evaluation approach aims at answering the question as to whether the models are capable to correctly classify a known location from a known device. 
In the data, we take the 10-minute readings collected from each device and divide it into ten 1-minute segments. We then proceed to a 70/30 cross evaluation by randomly selecting 3 segments for the testing set and the remaining 7 for the training set, i.e., there is no overlapping between training and test sets. We repeat this process and average over 30 iterations using the analysis methods detailed in Section~\ref{sec:feature_models}.
Under this evaluation criterion, we assume that we have (previous) data from the device being tested at the location from which the new observation is tested. 

\subsubsection{Leave-a-Place-Out.}
This evaluation approach aims at examining the location-type prediction performance from magnetic field readings belonging to locations that have not been seen by the model but belonging to the existing set of location-types. For instance, in this approach, we aim to classify a magnetic field reading that belongs to a train station but in a station different to the ones present in the training set. With this approach, we want to investigate whether the need to map each location can be eliminated and, thus, the location-type can be determined based on its characteristics.
In order to carry out this type of evaluation, we remove from each location-type a single distinct location at a time. We use the removed location as the test set and all the readings from the remaining locations in the training set. 
This process is repeated until each location is assigned once to the testing set.

\subsubsection{Leave-a-Device-Out.}
In this final evaluation approach, we are interested in determining whether the models can predict the location-type from the reading of a new device (\ie an unknown user). 
To carry out this evaluation, we select all readings from one device and use them in the testing set. The training set is then composed of all readings of the remaining devices. We use cross validation to evaluate the performance of the method. 

\section{Dataset}
\label{dataset}

Our data collection was carried out in three stages: (i) application development, (ii) data collection, and (iii) analysis and classification.

\subsection{Data Collection}
Our hypothesis in this project is that similar locations will be characterized by similar magnetic fields. We are attempting to determine whether a fingerprint exists for a particular location-type and use it to predict the class of unknown locations. This task requires on-site data collection at a number of locations across the city. The measurement devices must therefore be mobile and, if possible, able to be crowdsourced. For their sensors and ubiquity, we use cell phones to collect magnetic field readings at each location. 

We developed two applications, for Android and iPhone, and used 14 devices to collect measurements (6 Android phones \hl{running Android 7.1.2}, 8 iPhones \hl{running iOS 12}). Both applications collect the three dimensional magnetic field readings along with location, linear acceleration, user activity (for Android phones), and phone identifiers. In the evaluation (\ie testing) of each model we only considered the five devices for which we had data at all locations. 

\subsection{Description of the Dataset}


The final dataset consists of 550 magnetic-field measurements: we collected readings from 5 off-the-shelf smartphones at 110 different locations (10 location-types and 11 locations per type allowing us to train on at least 10 locations for each location-type).
Each magnetic reading contains at least 10 minutes of data sampled at 1~Hz. In our case, we are not interested in reconstructing the function and sampling at this rate allows us to record changes without overtaxing the resources of the phone. 
We designed a collection methodology such that the data that was collected mimic realistic life situations.
Volunteers were instructed to hold one phone while walking around the location as they would normally do for a period of 10 minutes. The samples were taken at different times of the day over a period of three months by different volunteers in an attempt to capture the inherent characteristics of a location rather than any bias introduced by the volunteers. 

The \textit{types} of places were chosen to fit three groups: indoor environments (\ie halls, gyms, laundromats, and coffee shops); outdoor environments (\ie parks and bridges); and transportation hubs (\ie subway stations, long-distance train stations, parking lots, bus stops). 
Locations, excluding the parks, are contained in an area of 9.7~mi$^2$. In any case, parks are contained within the greater metropolitan area of the city taken into consideration in this study.

\begin{figure*}[t]
    \begin{subfigure}[t]{0.5\textwidth}
        \centering
        \includegraphics[height=2in]{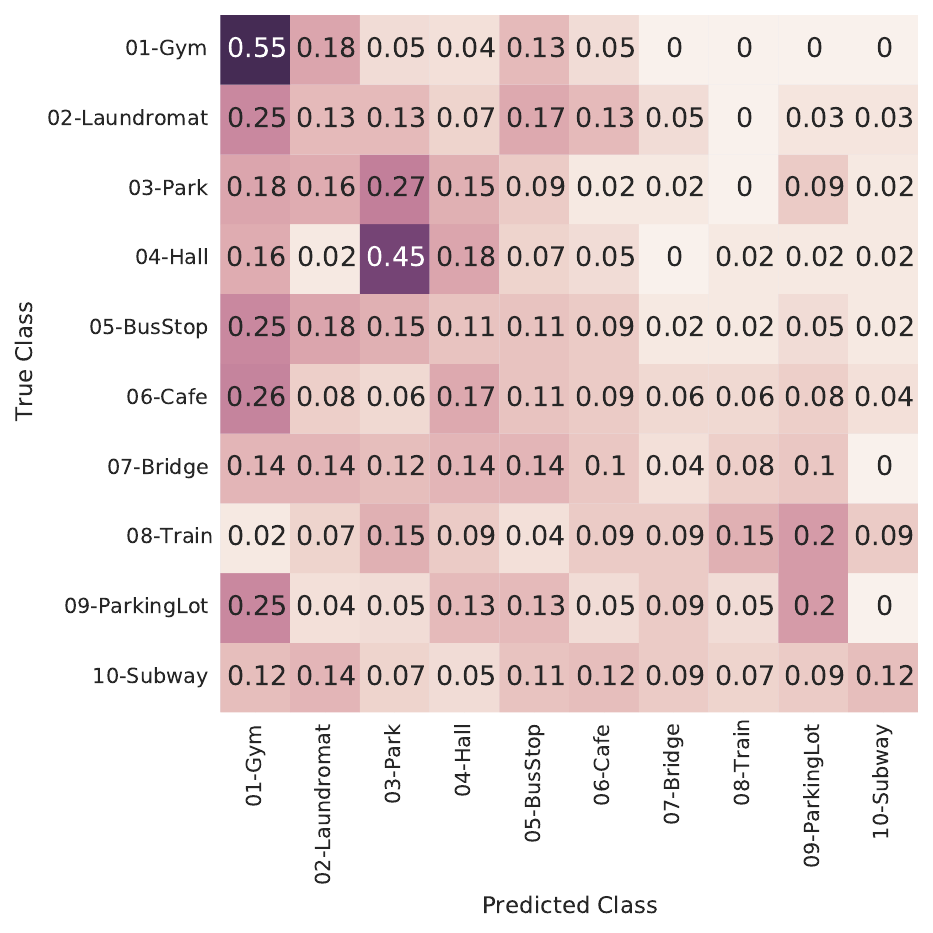}
        \caption{Full Signal Matching}
    \end{subfigure}%
    ~
    \begin{subfigure}[t]{0.5\textwidth}
        \centering
        \includegraphics[height=2in]{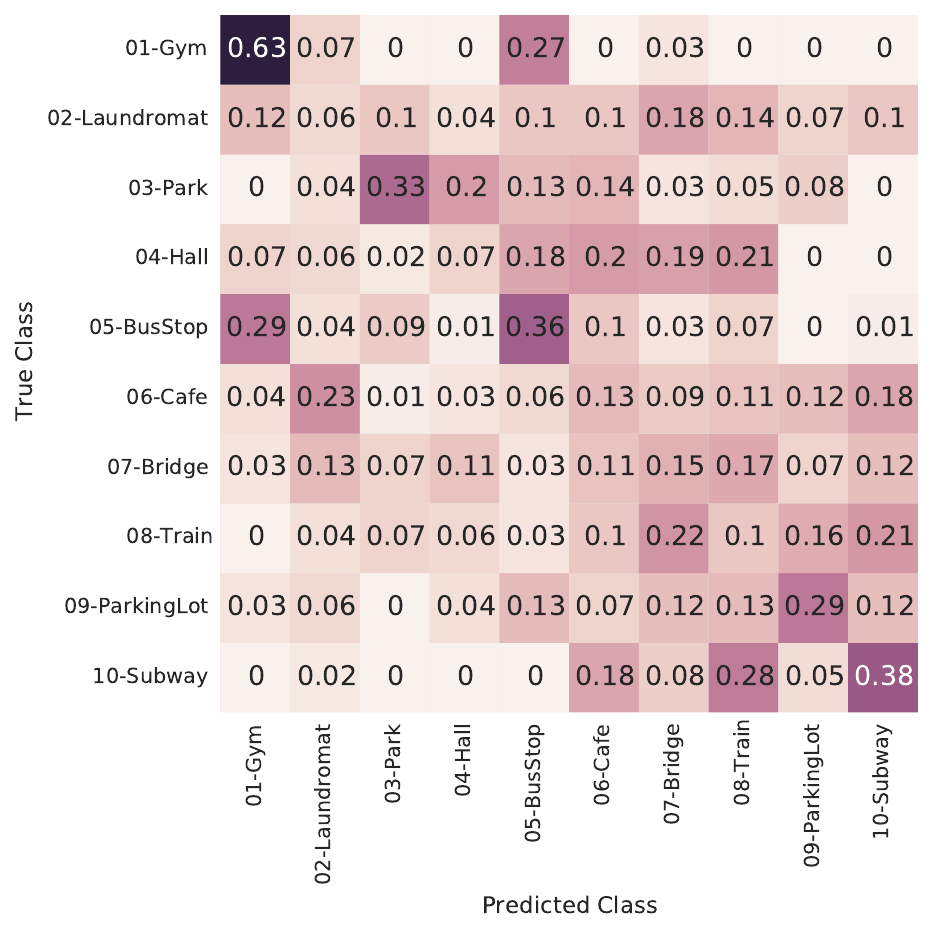}
        \caption{Statistical Descriptors}
    \end{subfigure}%
    ~
    \linebreak
    \begin{subfigure}[t]{0.5\textwidth}
        \centering
        \includegraphics[height=2in]{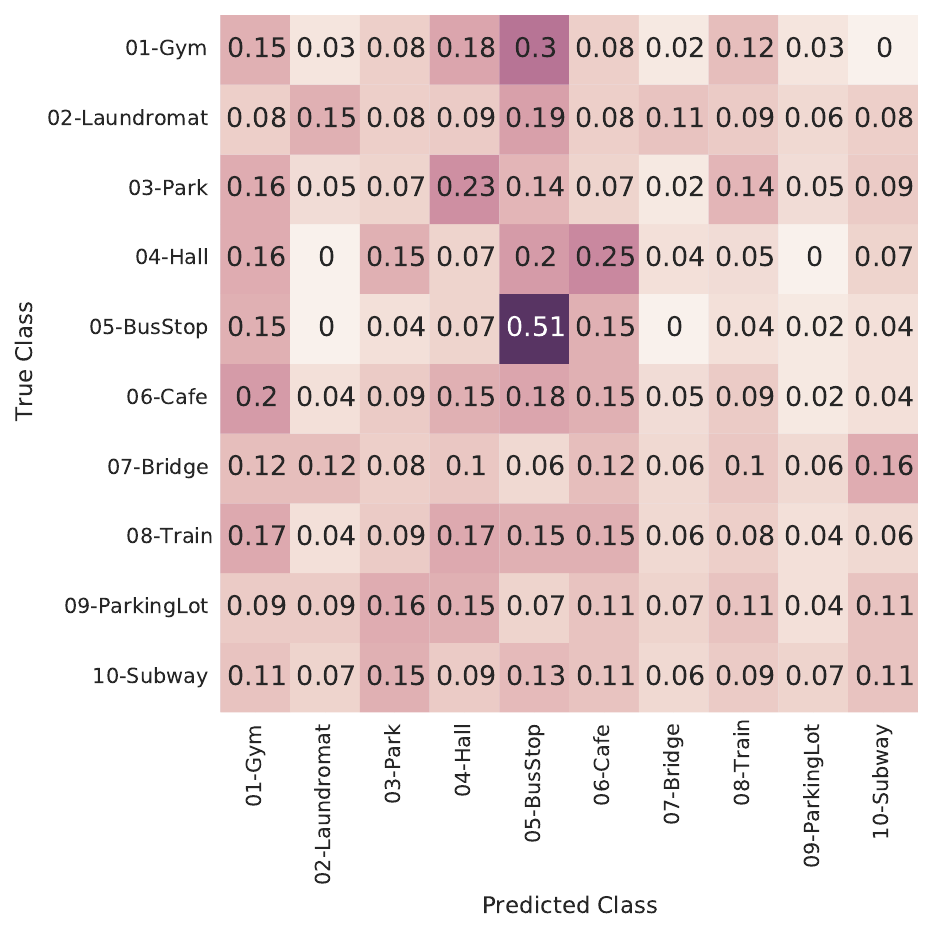}
        \caption{Automated Features}
    \end{subfigure}%
    ~
    \begin{subfigure}[t]{0.5\textwidth}
        \centering
        \includegraphics[height=2in]{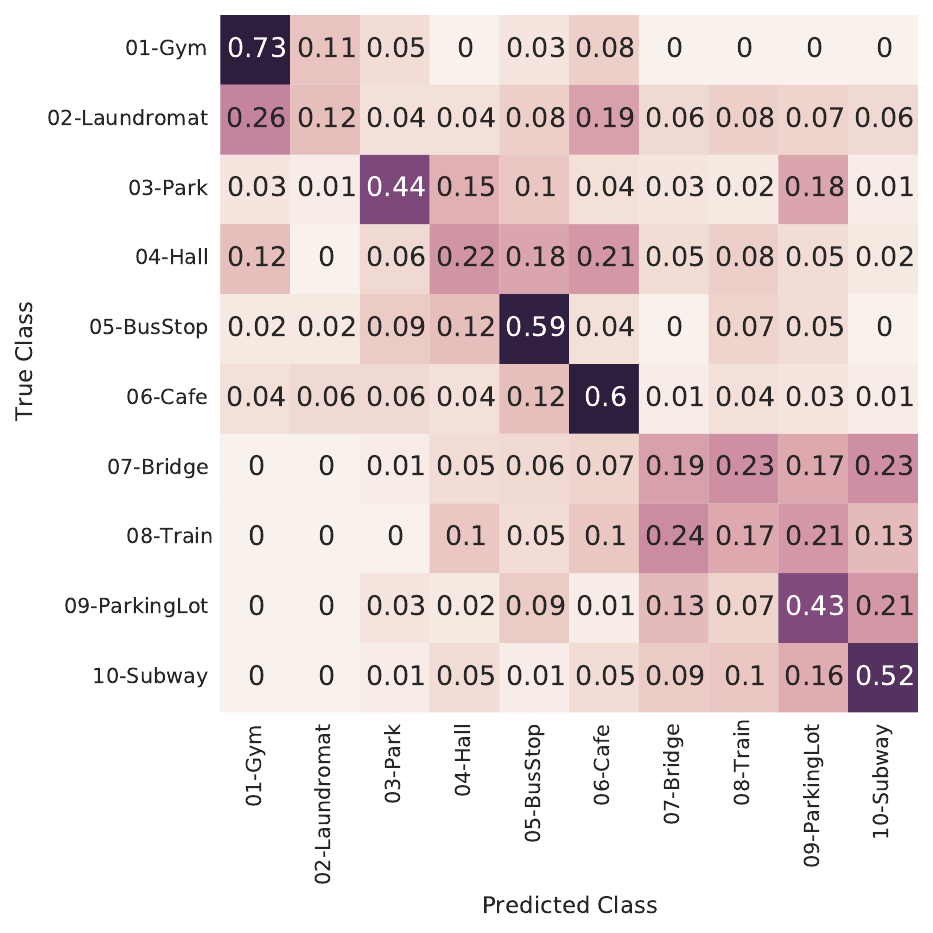}
        \caption{Shapelets}
    \end{subfigure}%

    \caption{Confusion matrices for the methods evaluated for the leave-a-place-out scenario for the best classifier/configuration.} 
    \label{Fig3}

\end{figure*}

\section{Results}

\label{section:Results}

In this section we present the evaluation of the four time-series classification methods, which we discussed in Section~\ref{sec:feature_models}, constructed to infer the location-types using magnetic field readings. As discussed in Section~\ref{sec:evaluation_criteria}, we evaluate each method using three criteria: (\textit{i}) all places, all devices, (\textit{ii}) leave-a-place-out, and (\textit{iii}) leave-a-device-out. 


\subsection{All Places, All Devices Evaluation} 

This evaluation corresponds to a scenario where the classifier has been trained with data originating from the device being tested at the location in question (\ie we have labeled data for all devices at every location). 
The separation between training and testing occurs over the temporal component of each observation. 
We \hl{converted} each observation into ten 1 minute segments and selected 3 random segments for the testing set. The remaining 7 segments become the training set. We \hl{repeated} this process 30 times to reduce the impact of the random selection in the results. 
This evaluation criterion resulted in 100\% accuracy for location-type identification from all devices. 

In this scenario, the attack requires having labeled data from all devices at each target location (\ie building a map per device with all locations). In a deployed system this would require having real-time labeled data to use as the basis of the map. Because of the pervasiveness of mobile devices this could be possible however, this would be the equivalent of the brute-force approach. In the following sections we present two scenarios where we explore  how to generalize the method for new devices and places.

\subsection{Leave-a-Place-Out Evaluation}
\label{LocationType:Place}

In this evaluation scenario, we \hl{used} the data of a single location (from all devices) as the test set and the data for the remaining 109 places as train set. We \hl{repeated} this until we use the data for each of the 110 places for testing, then \hl{aggregated} the results of all iterations to determine the overall performance. 


\hl{In order to better understand the performance of all methods, we present the confusion matrices for the four methods under consideration in Figures~\ref{Fig3}(a)-(d).}  In each matrix, the average of the diagonal values corresponds to the overall accuracy of the indicated classifier. In particular, we present the average accuracy for each classifier constructed for the feature extraction methods and evaluated using leave-a-place-out evaluation criterion. Overall, our results show that the classifier constructed with shapelets outperforms all others. 

The classifiers based on statistical descriptors and shapelets are optimized using three algorithms (\ie $k$NN, RF, and XGB). Our results show that both RF and XGB achieve the highest accuracy (with a negligible difference between them), whereas $k$NN has the worst performance. 
On the other hand, full signal matching classification \hl{was} optimized through four different distance metrics (\ie DTW, Euclidean, Cosine, and Bhattacharya distances). We \hl{observe} that DTW is the measure that performs best. Finally, we \hl{optimized} automated feature classification for different number of convolution layers (\ie 2, 3, and 4 layers with 4, 8, and 16 filters, respectively). The results for this optimization show that the model with 3 and 4 convolution layers achieve the best accuracy. \hl{We report these in Table \ref{table:LPO_combined}.}

We further \hl{investigated} whether the two best methods (\ie statistical descriptors and shapelets) extract and exploit different information from the data. To this end, we \hl{constructed} a new method that combines both of these feature sets as input. If both methods extract the same information, then the accuracy of the combined model should not improve. We \hl{compared} the results of the new combined feature classifier against statistical descriptors and shapelets used as baselines. \hl{ The results in Figure \ref{Fig4} shows that there it is possible to observe a performance improvement.} We \hl{constructed} the new combined feature set method by using the $k$NN, RF, and XGB classifiers. \hl{We report these results in Table \ref{LPO:combined-table}.}

\begin{table*}[t]
    \centering
    \caption{The performance of all classifiers for the leave-a-place-out evaluation.}
    \resizebox{.99\linewidth}{!}{%
    \begin{tabular}{|l|l||l|l||l|l||l|l|}
        \hline
        \multicolumn{2}{|c||}{ \textbf{(a)Full Signal Matching} } & \multicolumn{2}{c||}{ \textbf{(b)Statistical Descriptors} } & \multicolumn{2}{c||}{ \textbf{(c)Automated Features} }& \multicolumn{2}{c|}{ \textbf{(d)Shapelet} }  \Tstrut\Bstrut\\\hline
        \textit{Distance measure} & \textit{Accuracy} & \textit{Classifiers} & \textit{Accuracy} & \textit{CNN Layers} & \textit{Accuracy} & \textit{Classifiers} & \textit{Accuracy} \Tstrut\Bstrut\\\hline
        
        DTW & \textbf{0.1872} & $k$NN & 0.1170 & 2 & 0.0901 & $k$NN & 0.2690 \Tstrut\Bstrut\\
        Euclidean & 0.1750 & RF & 0.2040 & 3 & \textbf{0.0907} & RF & \textbf{0.4045} \\
        Cosine & 0.1713 & XGB & \textbf{0.2100} & 4 & 0.0902 & XGB & 0.3920 \\
        Bhattacharyya & 0.1005 & & & & & & \Bstrut\\\hline
    \end{tabular}}
    \label{table:LPO_combined}
\end{table*}

\begin{table*}[t]
\begin{minipage}[h]{0.3\linewidth}
\centering
    \caption{Leave-a-Place-Out Classification}
    \resizebox{\linewidth}{!}{%
        \begin{tabular}{|l|l|}
            \hline
            \multicolumn{2}{|c|}{\textbf{Combined Feature Set} }
            \Tstrut\Bstrut\\\hline
            \textit{Classifiers} & \textit{Accuracy} \Tstrut\Bstrut\\\hline
             $k$NN & 0.1860 \Tstrut\Bstrut\\
              RF  & \textbf{0.3907} \Tstrut\Bstrut\\
             XGB  & 0.3442  \Tstrut\Bstrut\\
            \hline
        \end{tabular}}
    \label{LPO:combined-table}
\end{minipage}\hfill
\vspace{0pt}
\begin{minipage}[h]{0.6\linewidth}
\centering
\includegraphics[height=2in]{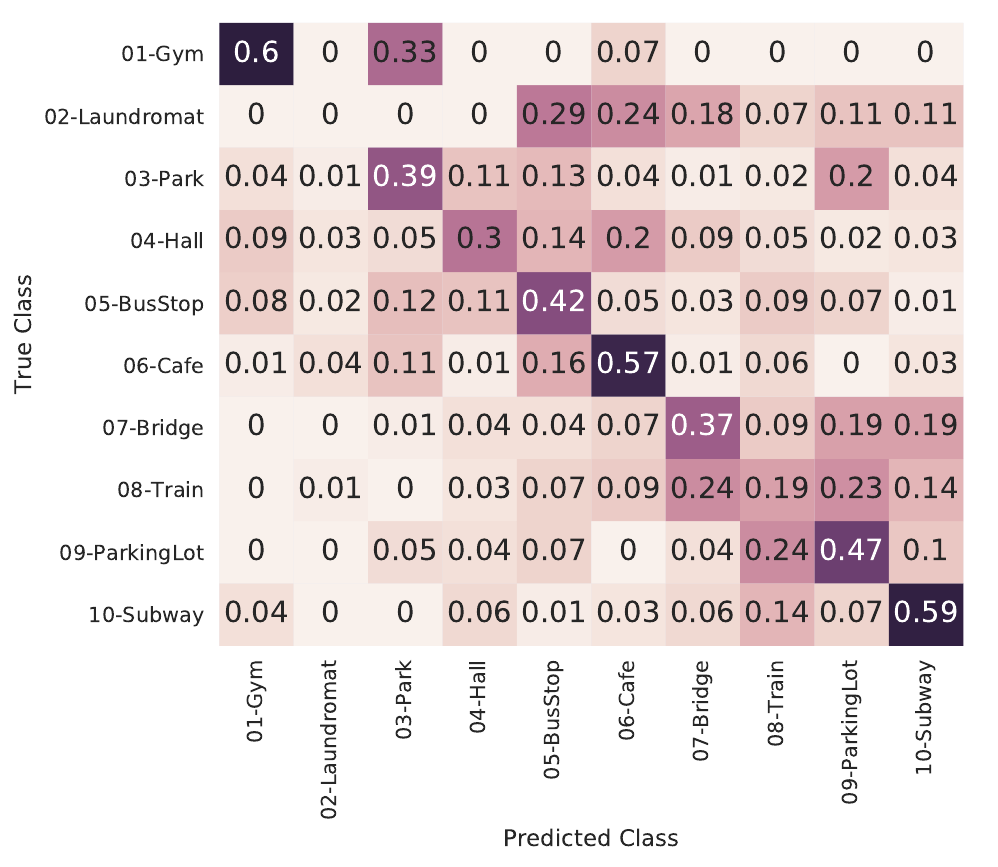}
\captionof{figure}{Confusion Matrix for the best configuration: Leave-a-Place-Out.}
\label{Fig4}
\end{minipage}
\end{table*}



\subsection{Leave-a-Device-Out Evaluation}
\label{LocationType:Device}

The readings collected by mobile devices have some error due to the quality of the sensors, their age, and usage~\cite{baldini2017identification}. Each device has the software necessary to compensate for the error and provide measurements close to the actual value. As such, this error-correction \hl{enabled} us to crowd-source the data collection in order to process and extract more robust descriptors for each \textit{location-type} as we would be able to observe the same places under different environmental settings. We \hl{evaluated} our methods using a leave-one-device-out scenario as discussed in Section~\ref{sec:evaluation_criteria}.

\begin{figure*}[t]
    \begin{subfigure}[t]{0.5\textwidth}
        \centering
        \includegraphics[height=2in]{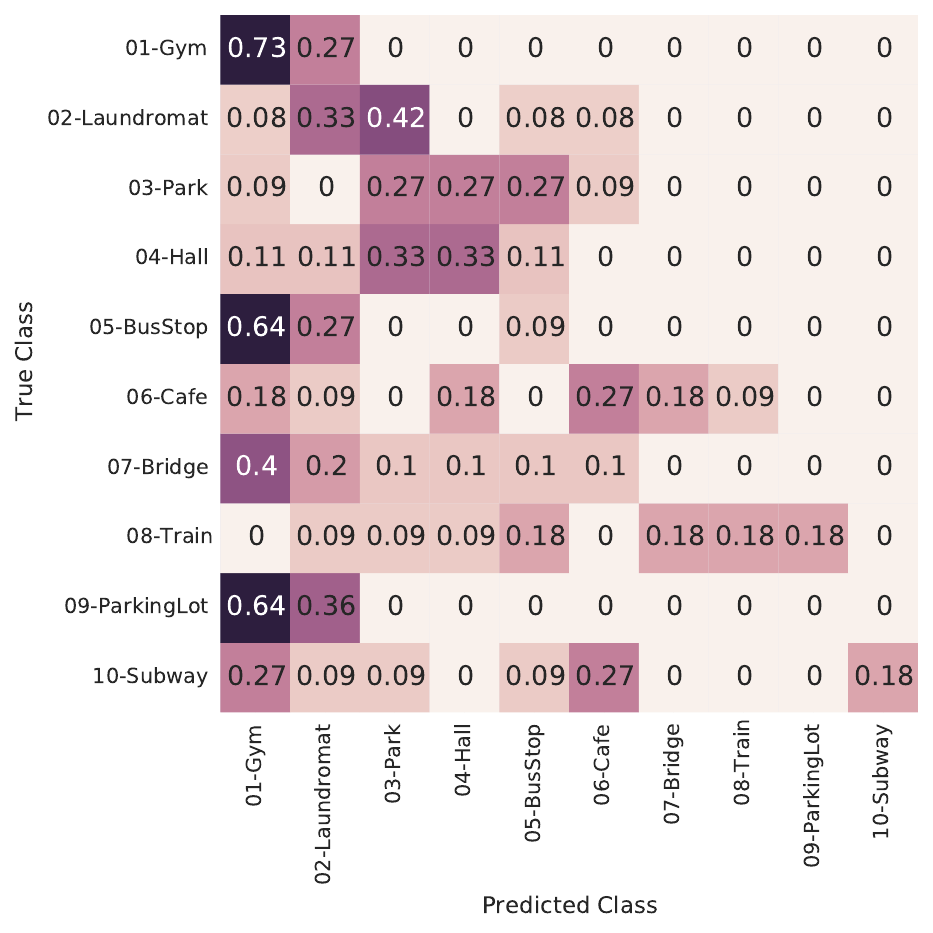}
        \caption{Full Signal Matching}
    \end{subfigure}%
    ~
        \begin{subfigure}[t]{0.5\textwidth}
        \centering
        \includegraphics[height=2in]{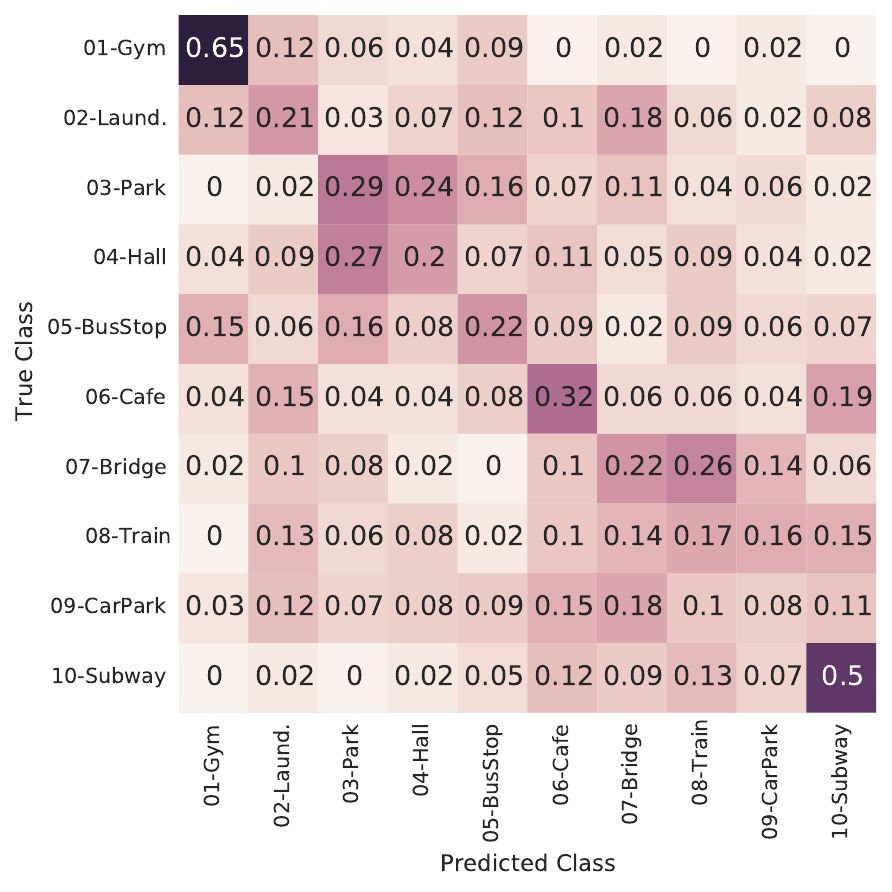}
        \caption{Statistical Descriptors}
    \end{subfigure}%
    ~
    \linebreak
        \begin{subfigure}[t]{0.5\textwidth}
        \centering
        \includegraphics[height=2in]{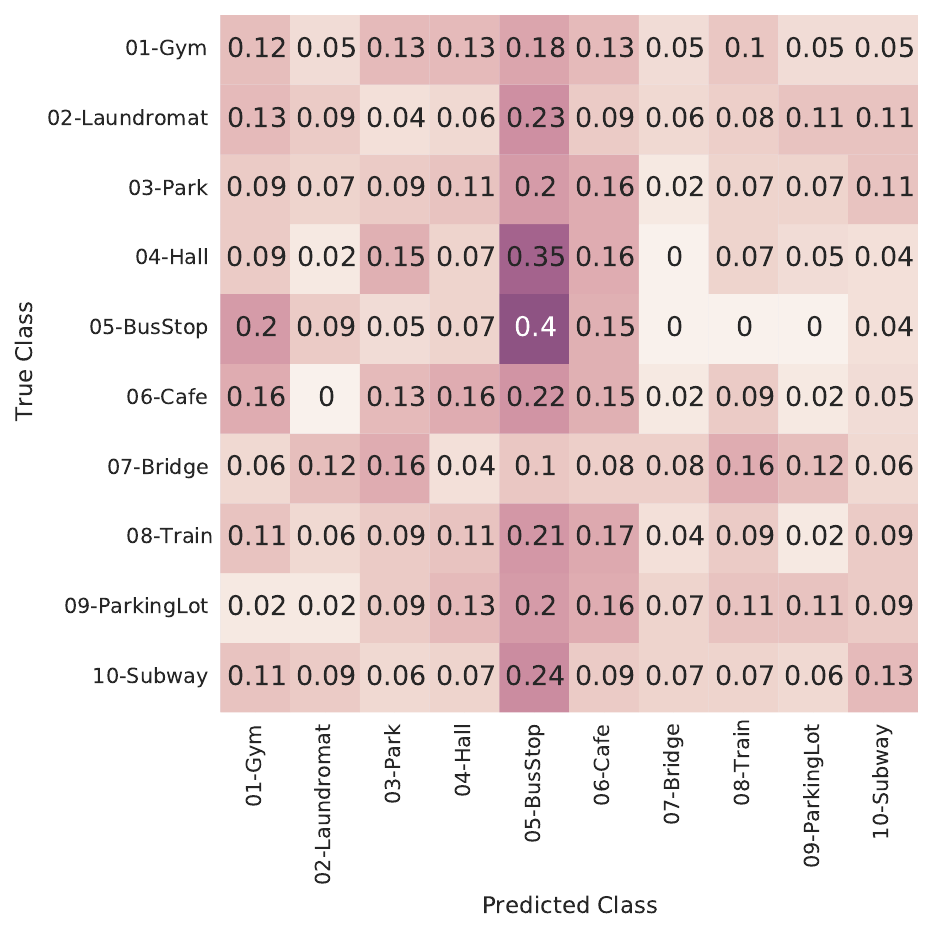}
        \caption{Automated Features}
    \end{subfigure}%
    ~
    \begin{subfigure}[t]{0.5\textwidth}
        \centering
        \includegraphics[height=2in]{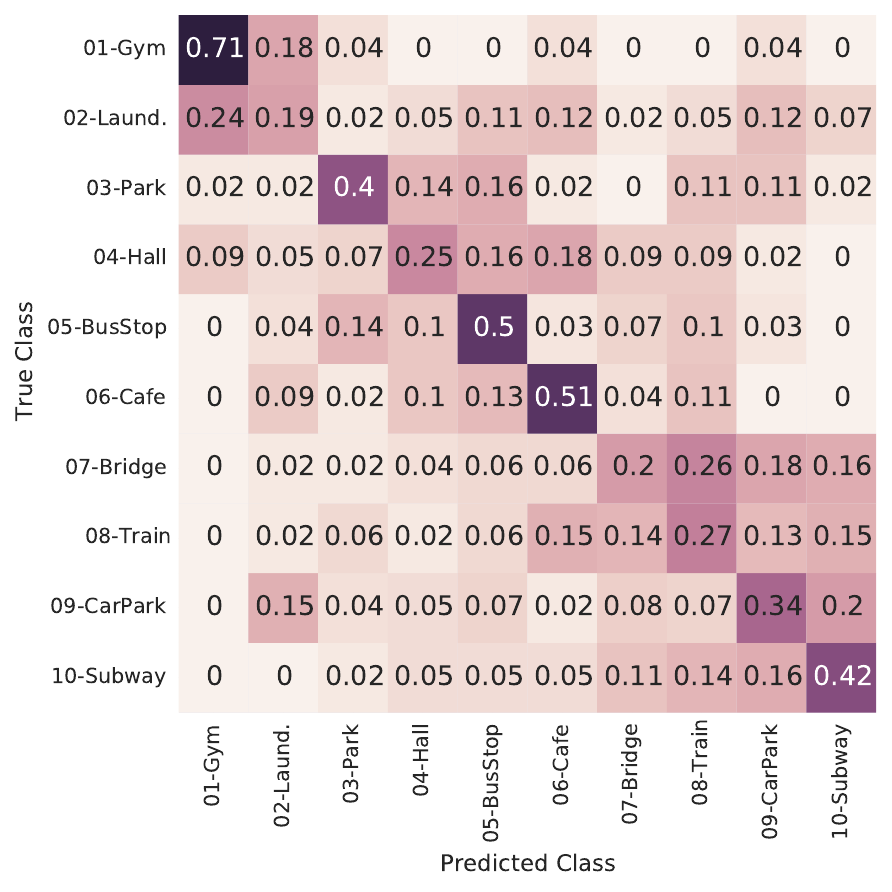}
        \caption{Shapelets}
    \end{subfigure}%
    ~
    \caption{Confusion matrices for the methods evaluated for the leave-a-device-out scenario for the best classifier/configuration.} 
    \label{Fig5}
\end{figure*}

%
%

\hl{In Figures~\ref{Fig5}(a)-(d) we present the confusion matrices} for the best performing models for each four methods presented in Section~\ref{sec:feature_models}.
%
Similar to the leave-a-place-out scenario, we \hl{optimized} the methods based on statistical descriptors and shapelets using the same three classifiers (\ie $k$NN, RF, and XGB). Our results \hl{remained} consistent: both RF and XGB achieve the highest accuracy (with a negligible difference between them). \hl{These results are presented in Table \ref{LDO:combined-table}.}
Full signal matching classification \hl{was} optimized through four different distance metrics (\ie DTW, Euclidean, Cosine, and Bhattacharya distances); we \hl{found} that the method that works best uses Euclidean distance as its metric. Finally, we \hl{optimized} the automated features classifier for different number of convolution layers (\ie 2, 3, and 4 layers with 4, 8, and 16 filters respectively). The results are again consistent with the previous analysis: the model with 3 and 4 convolution layers achieve the best accuracy. \hl{These results are reported in Table \ref{table:LDO_combined}.}

We \hl{investigated} whether the two best methods (\ie statistical descriptors and shapelets) extract and exploit different information from the magnetic field readings. 
We constructed a new classifier that combines both feature sets as input and compares the results against statistical descriptors and shapelets based methods (used as baselines). The evaluation results show that the  classifier based on the combined features achieves an accuracy improvement of 7\% as compared to the best of the previous methods. The increase in performance by the combined method can be attributed to the fact that the statistical descriptors exclude the outlying values from a dataset, whereas they are included by the shapelets in the temporal patterns~\cite{fu2011review}. This shows the consistency of the two methods (\ie shapelets and statistical descriptors) for extracting complementary information from the time-series data. \hl{We present these results in Figure \ref{Fig6}.}
%


\begin{table*}[t]
\begin{minipage}[h]{0.3\linewidth}
\centering
    \caption{Leave-a-Device-Out Classification}
    \resizebox{\linewidth}{!}{%
        \begin{tabular}{|l|l|}
            \hline
            \multicolumn{2}{|c|}
            {\textbf{Combined Feature Set} }
            \Tstrut\Bstrut\\\hline
            \textit{Classifiers} & \textit{Accuracy} \Tstrut\Bstrut\\\hline
             $k$NN & 0.1776 \Tstrut\Bstrut\\
              RF  & \textbf{0.4256} \Tstrut\Bstrut\\
             XGB  & 0.4200  \Tstrut\Bstrut\\
            \hline
        \end{tabular}}
    \label{LDO:combined-table}
\end{minipage}\hfill
\vspace{0pt}
\begin{minipage}[h]{0.6\linewidth}
\centering
\includegraphics[height=2in]{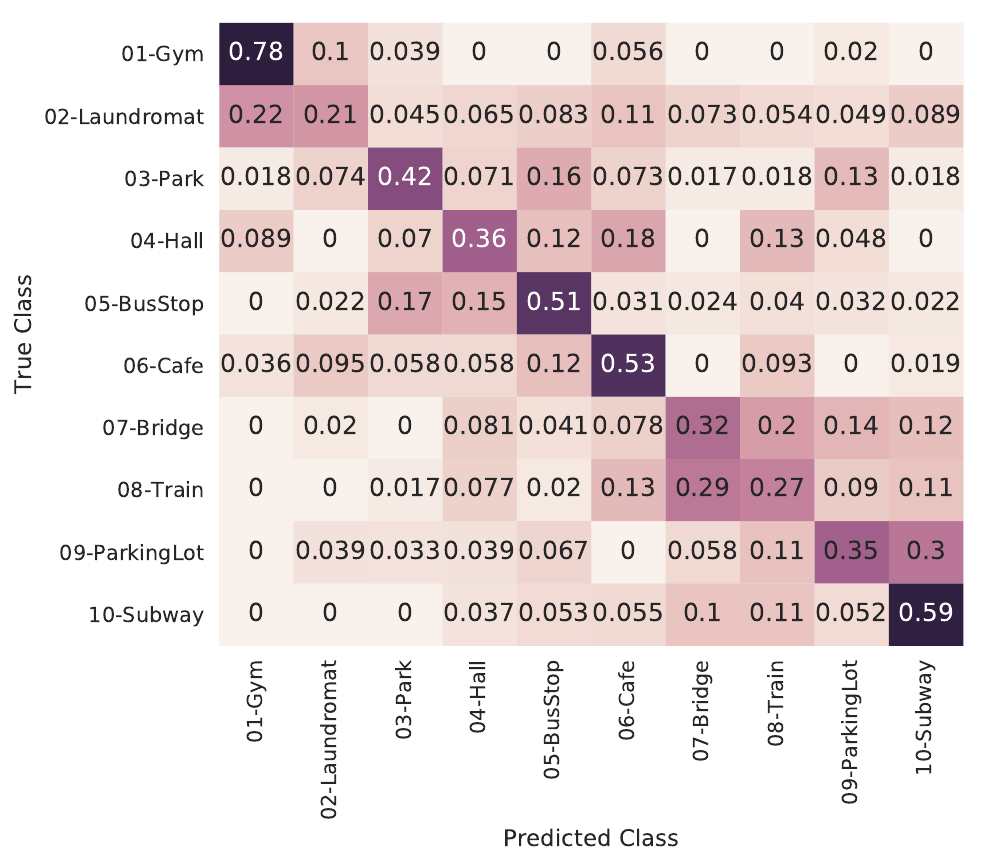}
\captionof{figure}{Classification for the combined feature set for the two most accurate techniques: shapelets and statistical descriptors}
\label{Fig6}
\end{minipage}
\end{table*}

\begin{table*}[t]
    \centering
    \caption{Performance of all classifiers for the leave-a-device-out evaluation.}
        \resizebox{.99\linewidth}{!}{%
        \begin{tabular}{|l|l||l|l||l|l||l|l|}
        \hline
        \multicolumn{2}{|c||}{ \textbf{(a)Full Signal Matching} } & \multicolumn{2}{c||}{ \textbf{(b)Statistical Descriptors} } & \multicolumn{2}{c||}{ \textbf{(c)Automated Features} } & \multicolumn{2}{c|}{ \textbf{(d)Shapelets} }\Tstrut\Bstrut\\\hline
        \textit{Distance measure} & \textit{Accuracy} & \textit{Classifiers} & \textit{Accuracy} & \textit{CNN Layers} & \textit{Accuracy} & \textit{Classifiers} & \textit{Accuracy} \Tstrut\Bstrut\\\hline
        
        DTW & 0.1852 & $k$NN & 0.1776 & 2 & 0.1025 & $k$NN & 0.2495 \Tstrut\Bstrut\\
        Euclidean &\textbf{0.2407} & RF & \textbf{0.2998} & 3 & \textbf{0.1291} & RF & 0.3552  \\
        Cosine & 0.1481 & XGB & 0.2926 & 4 & 0.1274 & XGB & \textbf{0.3593} \\
        Bhattacharyya & 0.1204 & & & & & & \Bstrut\\\hline
    \end{tabular}}
    \label{table:LDO_combined}
\end{table*}

\begin{table}[t]
\centering
    \caption{Leave-a-Device-Out: Frequency domain.}
    \resizebox{0.5\linewidth}{!}{%
        \begin{tabular}{|l|c|c|c|}
            \hline
            \multicolumn{4}{|c|}{\textbf{Frequency Domain}}
            \Tstrut\Bstrut\\\hline
            &\textit{Shapelets} & \textit{Statistical Descriptors} & \textit{Combined} \Tstrut\Bstrut\\\hline
             $k$NN & 0.1750 & 0.1701 & 0.1640 \Tstrut\Bstrut\\
             RF & \textbf{0.2030} & 0.3095 & 0.2634\Tstrut\Bstrut\\
             XGB  & 0.1675 & \textbf{0.3147} & \textbf{0.2790}  \Tstrut\Bstrut\\
            \hline
        \end{tabular}}
    \label{LDO:freq-table}
\end{table}


\hl{
\section{Threats to Validity}
The proposed method depends on the possibility of accessing the magnetometer through zero-permission or through involuntary access to it through an app. We also assume that the magnetometer’s refresh rate is high enough to allow shapelet extraction. This might change in the future and/or might be different in the future. It is worth noting that, for example, rate limitation for a variety of position sensors has been introduced in Android 12 \cite{sensor-overview}.
In addition, different places might be characterized by similar geo-magnetic profiles. Our tests have been limited in terms of variety of places and we cannot exclude that this might happen in certain circumstances. More in general, it might happen that the environment is highly noisy. This might also affect the applicability of the proposed method in general.

Finally, in terms of resources, an application characterized by a disproportionate use of memory (used to store readings) and battery (for network transmission continuous observations) might look suspicious. In contrast, a system where shapelets are computed offline and transferred to the phone for classification will be practically unnoticeable to the user. Indeed, after training, classification in itself is a resource-efficient task. In our implementation, once computed, all shapelets (\ie the features) required approximately 50~kB of storage. 

}
\section{Summary of the Contributions}
\label{conclusion}

We show that magnetic field measurements can be leveraged to identify types of places with approximately 40\% accuracy. We infer location-types both from unseen places and unseen devices. To the best of our knowledge, we are the first to present such an approach. Most of the literature in localization bases location on maps: either by matching the exact values at one location or calculating the distance from some known position. We instead focus on the physical features of the environment: their magnetic signatures.
In order to perform the experimental evaluation of the proposed attack, we performed an in-the-wild measurement study. We collected over 91 hours of data from 10 location-types across a major metropolitan area and present our results from a set of 110 distinct locations. The collection times at each location varied in terms of time of day, day of the week, and, for some locations, the readings between phones differ by up to a month. We expect that by allowing such variation, we mitigated the impact of external influences on the generalizability of our results.

Overall, this paper shows that environmental characteristics can be leveraged to infer the \textit{location-type} achieving coarse-grained localization. We have presented two scenarios, one where the test location has never been visited and another where the phone that is collecting the measurements has never been seen. We believe that this constitutes a substantial contribution as the ability to determine coarse-grained location information without the need to map all the locations of interest. Moreover, we find that, as a passive, zero-permission attack it can be used to track users, inferring their routines.

\begin{credits}
\subsubsection{\ackname} Mirco Musolesi was supported by the Italian Ministry of University and Research (MUR) through the project PRIN 2022 “Machine-learning based control of complex multi-agent systems for search and rescue operations in natural disasters (MENTOR)” funded by the European Union - NextGenerationEU.

\end{credits}
%
%
%
\bibliographystyle{splncs04}
\bibliography{sample-base}

@MISC{androidPermissionLevel,
  year={2022},
  author = {Android},
  title = {{Location Permissions}},
  howpublished = {\url{https://developer.android.com/reference/android/Manifest.permission}},
  note = {Accessed: 2024-04-20}
}

@MISC{WindowsPermission,
  year={2018},
  author = {Microsoft},
  title = {{Windows Devices: Geolocation}},
  howpublished = {\url{https://docs.microsoft.com/en-us/uwp/api/Windows.Devices.Geolocation}},
  note = {Accessed: 2024-04-20}
}

@MISC{ApplePermission,
  year={2018},
  author = {Apple},
  title = {{Apple Developer: Determining the Availability of Location Services}},
  howpublished = {\url{https://developer.apple.com/documentation/corelocation/determining_the_availability_of_location_services}},
  note = {Accessed: 2024-04-20}
}

@ARTICLE{deMontjoye2013unique,
  title={{Unique in the Crowd: The Privacy Bounds of Human Mobility}},
  author={De Montjoye, Yves-Alexandre and Hidalgo, C{\'e}sar A and Verleysen, Michel and Blondel, Vincent D},
  journal={Scientific reports},
  volume={3},
  pages={1376},
  year={2013},
  publisher={Nature Publishing Group}
}

@INPROCEEDINGS{li2008mining,
  title={{Mining User Similarity based on Location History}},
  author={Li, Quannan and Zheng, Yu and Xie, Xing and Chen, Yukun and Liu, Wenyu and Ma, Wei-Ying},
  booktitle={ACM SIGSPATIAL},
  pages={34},
  year={2008},
  month=nov,
  address={Irvine, CA, USA},
  publisher={ACM}
}

@INPROCEEDINGS{kautz2004learning,
  title={{Learning and Inferring Transportation Routines}},
  author={Lin Liao and Donald J. Patterson and Dieter Fox and Henry Kautz},
  booktitle={AAAI},
  month=jul,
  year={2004},
  address={San Jose, CA, USA}
}

@INPROCEEDINGS{chen2012fmloc,
 author = {Chen, Yin and Lymberopoulos, Dimitrios and Liu, Jie and Priyantha, Bodhi},
 title = {{FM-based Indoor Localization}},
 booktitle = {ACM MobiSys},
 year = {2012},
 month = jun,
 address = {Low Wood Bay, UK}
}

@inproceedings{michalevsky2015powerspy,
  title={{PowerSpy: Location Tracking Using Mobile Device Power Analysis}},
  author={Michalevsky, Yan and Schulman, Aaron and Veerapandian, Gunaa Arumugam and Boneh, Dan and Nakibly, Gabi},
  address={Washington, D.C., USA},
  booktitle={USENIX Conference on Security Symposium},
  month=aug,
  year=2015,
  pages={785-800}
}

@ARTICLE{Wang2010survey,
  author={Wang, Jing and Ghosh, R. K.and Das, Sajal K.},
  title={{A Survey on Sensor Localization}},
  journal={Journal of Control Theory and Applications},
  year={2010},
  month=feb,
  volume={8},
  number={1},
  pages={2-11}
}

@INPROCEEDINGS{zhou2013identity,
 author = {Zhou, Xiaoyong and Demetriou, Soteris and He, Dongjing and Naveed, Muhammad and Pan, Xiaorui and Wang, XiaoFeng and Gunter, Carl A. and Nahrstedt, Klara},
 title = {{Identity, Location, Disease and More: Inferring Your Secrets from Android Public Resources}},
 booktitle = {ACM CCS},
 year = {2013},
 month = nov,
 address = {Berlin, Germany}
}

@ARTICLE{vo16outdoorSurvey, 
  author={Q. D. Vo and P. De},
  journal={IEEE Communications Surveys Tutorials}, 
  title={A Survey of Fingerprint-Based Outdoor Localization}, 
  year={2016}, 
  volume={18}, 
  number={1}, 
  pages={491-506}
}

@INPROCEEDINGS{chung11indoor,
 author = {Chung, Jaewoo and Donahoe, Matt and Schmandt, Chris and Kim, Ig-Jae and Razavai, Pedram and Wiseman, Micaela},
 title = {{Indoor Location Sensing Using Geo-magnetism}},
 booktitle = {ACM MobiSys},
 address={Bethesda, MD, USA},
 month=jun,
 year = {2011}
}

@ARTICLE{gozick11magnetic, 
  author={B. Gozick and K. P. Subbu and R. Dantu and T. Maeshiro}, 
  journal={IEEE Transactions on Instrumentation and Measurement}, 
  title={{Magnetic Maps for Indoor Navigation}}, 
  month=dec,
  year={2011}, 
  volume={60}, 
  number={12}, 
  pages={3883-3891}
}

@INPROCEEDINGS{wang12wardrive,
  author = {Wang, He and Sen, Souvik and Elgohary, Ahmed and Farid, Moustafa and Youssef, Moustafa and Choudhury, Romit Roy},
  title = {{No Need to War-drive: Unsupervised Indoor Localization}},
  booktitle = {ACM MobiSys},
  address={Low Wood Bay, UK},
  month=jun,
  year = {2012}
}

@article{bagnall2017great,
  title={{The Great Time Series Classification Bake Off: A Review and Experimental Evaluation of Recent Algorithmic Advances}},
  author={Bagnall, Anthony and Lines, Jason and Bostrom, Aaron and Large, James and Keogh, Eamonn},
  journal={Data Mining and Knowledge Discovery},
  volume={31},
  number={3},
  pages={606--660},
  year={2017},
  publisher={Springer}
}

@article{aghabozorgi2015clustering,
    title = {{Time-series Clustering~--~ A Decade Review}},
    journal = {Information Systems},
    volume = {53},
    pages = {16 - 38},
    year = {2015},
    author = {Saeed Aghabozorgi and Ali Seyed Shirkhorshidi and Teh Ying Wah},
}

@article{haverinen09global,
  title = {{Global Indoor Self-localization Based on the Ambient Magnetic Field}},
  author = {Janne Haverinen and Anssi Kemppainen},
  journal = {Robotics and Autonomous Systems},
  volume = {57},
  number = {10},
  pages = {1028-1035},
  month=oct,
  year = {2009}
}

@BOOK{thrun2005probabilistic,
  title={{Probabilistic Robotics}},
  author={Thrun, Sebastian and Burgard, Wolfram and Fox, Dieter},
  year={2005},
  publisher={MIT Press}
}

@ARTICLE{wu13will, 
  author={C. Wu and Z. Yang and Y. Liu and W. Xi}, 
  journal={IEEE Transactions on Parallel and Distributed Systems}, 
  title={{WILL: Wireless Indoor Localization without Site Survey}}, 
  month=apr,
  year={2013}, 
  volume={24}, 
  number={4}, 
  pages={839-848}
}

@INPROCEEDINGS{chintalapudi2010nopain,
 author = {Chintalapudi, Krishna and Padmanabha Iyer, Anand and Padmanabhan, Venkata N.},
 title = {{Indoor Localization Without the Pain}},
 booktitle = {ACM MobiCom},
 year = {2010},
 address = {Chicago, IL, USA},
 month= sep
}

@INPROCEEDINGS{yang09rss, 
  author={J. Yang and Y. Chen}, 
  title={{Indoor Localization Using Improved RSS-Based Lateration Methods}}, 
  booktitle={IEEE GLOBECOM}, 
  address={Honolulu, HI, USA},
  month=nov,
  year={2009}
}

@ARTICLE{chang10wifi, 
  author={N. Chang and R. Rashidzadeh and M. Ahmadi}, 
  journal={IEEE Transactions on Consumer Electronics}, 
  title={{Robust Indoor Positioning using Differential WiFi Access Points}}, 
  month=aug,
  year={2010}, 
  volume={56}, 
  number={3}, 
  pages={1860-1867}
}

@inproceedings{woodman08pedestrian,
 author = {Woodman, Oliver and Harle, Robert},
 title = {{Pedestrian Localisation for Indoor Environments}},
 booktitle = {ACM UbiComp},
 address={Seoul, Korea},
 month=sep,
 year = {2008}
}

@inproceedings{narain2016inferring,
 author = {S. Narain and T. D. Vo-Huu and K. Block and G. Noubir},
 title={{Inferring User Routes and Locations Using Zero-Permission Mobile Sensors}},
 booktitle = {IEEE S\&P},
 address={San Jose, CA, USA},
 month=may,
 year = {2016}
}

@inproceedings{block_magnetometer, author = {Block, Kenneth and Noubir, Guevara}, title = {{My Magnetometer Is Telling You Where I've Been? A Mobile Device Permissionless Location Attack}}, year = {2018}, isbn = {9781450357319}, publisher = {Association for Computing Machinery}, address = {New York, NY, USA}, booktitle = {WiSec}, pages = {260–270}, numpages = {11}, location = {Stockholm, Sweden}, }

@article{zhang2013senstrack,
  title={{SensTrack: Energy-Efficient Location Tracking with Smartphone Sensors}},
  author={Zhang, Lei and Liu, Jiangchuan and Jiang, Hongbo and Guan, Yong},
  journal={IEEE Sensors Journal},
  volume={13},
  number={10},
  pages={3775--3784},
  year={2013}
}

@inproceedings{han2012accomplice,
  title={{ACcomplice: Location Inference using Accelerometers on Smartphones}},
  author={Han, Jun and Owusu, Emmanuel and Nguyen, Le T and Perrig, Adrian and Zhang, Joy},
  booktitle={IEEE COMSNETS},
  year={2012},
  month=jan
}

@inproceedings{nawaz2014mining,
  title={{Mining Users' Significant Driving Routes with Low-Power Sensors}},
  author={Nawaz, Sarfraz and Mascolo, Cecilia},
  booktitle={ACM SenSys},
  address={Memphis, TN, USA},
  month=nov,
  year={2014}
}

@inproceedings{rai2012zee,
 author = {Rai, Anshul and Chintalapudi, Krishna Kant and Padmanabhan, Venkata N. and Sen, Rijurekha},
 title = {{Zee: Zero-effort Crowdsourcing for Indoor Localization}},
 booktitle = {ACM MobiCom},
 year = {2012},
 month=aug,
 address = {Istanbul, Turkey}
}

@INPROCEEDINGS{hadsell2006dimensionality,
  title={{Dimensionality Reduction by Learning an Invariant Mapping}},
  author={Hadsell, Raia and Chopra, Sumit and LeCun, Yann},
  booktitle={IEEE CVPR},
  year={2006},
  address={New York, NY, USA},
  month=jun
  }

@INPROCEEDINGS{zakaria2012clustering,
  title={{Clustering Time Series using Unsupervised Shapelets}},
  author={Zakaria, Jesin and Mueen, Abdullah and Keogh, Eamonn},
  booktitle={IEEE ICDM},
  address={Brussels, Belgium},
  month=dec,
  year={2012}
}

@INPROCEEDINGS{mueen2011logical,
  title={{Logical-Shapelets: An Expressive Primitive for Time Series Classification}},
  author={Mueen, Abdullah and Keogh, Eamonn and Young, Neal},
  booktitle={ACM KDD},
  address={ San Diego, CA, USA},
  month=aug,
  year={2011}
}

@article{fu2011review,
  title={{A Review on Time Series Data Mining}},
  author={Fu, Tak-Chung},
  journal={Engineering Applications of Artificial Intelligence},
  volume={24},
  number={1},
  pages={164--181},
  month=feb,
  year={2011},
  publisher={Elsevier}
}

@inproceedings{ye2009time,
  title={{Time Series Shapelets: A New Primitive for Data Mining}},
  author={Ye, Lexiang and Keogh, Eamonn},
  booktitle={ACM KDD},
  address={Paris, France},
  month=jun,
  year={2009}
}

@inproceedings{grabocka2014learning,
  title={{Learning Time Series Shapelets}},
  author={Grabocka, Josif and Schilling, Nicolas and Wistuba, Martin and Schmidt-Thieme, Lars},
  booktitle={ACM KDD},
  address={New York, NY, USA},
  month=aug,
  year={2014}
}

@article{hills2014classification,
  title={{Classification of Time Series by Shapelet Transformation}},
  author={Hills, Jon and Lines, Jason and Baranauskas, Edgaras and Mapp, James and Bagnall, Anthony},
  journal={Data Mining and Knowledge Discovery},
  volume={28},
  number={4},
  pages={851--881},
  month=jul,
  year={2014},
  publisher={Springer}
}

@ARTICLE{baldini2017identification,
  author = {Baldini, Gianmarco and Dimc, Franc and Kamnik, Roman and Steri, Gary and Giuliani, Raimondo and Gentile, Claudio},
  title = {{Identification of Mobile Phones using the Built-in Magnetometers Stimulated by Motion Patterns}},
  journal = {Sensors},
  year = {2017},
  volume = {17},
  pages = {783},
  number = {4},
  publisher = {Multidisciplinary Digital Publishing Institute}
}

@ARTICLE{pedregosa2011scikit,
  author = {Pedregosa, Fabian and Varoquaux, Ga{\"e}l and Gramfort, Alexandre and Michel, Vincent and Thirion, Bertrand and Grisel, Olivier and Blondel, Mathieu and Prettenhofer, Peter and Weiss, Ron and Dubourg, Vincent},
  title = {{Scikit-learn: Machine Learning in Python}},
  journal = {Journal of Machine Learning Research},
  number = oct,
  year = {2011},
  volume = {12},
  pages = {2825--2830}
}

@ARTICLE{breiman2001randomForest,
  author = {Breiman, Leo},
  title = {{Random Forests}},
  journal = {Machine Learning},
  month=oct,
  year = {2001},
  volume = {45},
  pages = {5--32},
  number = {1},
  publisher = {Springer}
}

@ARTICLE{cover1967knn,
  author = {T. Cover and P. Hart},
  title = {{Nearest Neighbor Pattern Classification}},
  journal = {IEEE Transactions on Information Theory},
  month = jan,
  year = {1967},
  volume = {13},
  pages = {21-27},
  number = {1}
}

@inproceedings{chen2016xgb,
 author = {Chen, Tianqi and Guestrin, Carlos},
 title = {{XGBoost: A Scalable Tree Boosting System}},
 booktitle = {ACM KDD},
 month = aug,
 year = {2016},
 address = {San Francisco, CA, USA}
}

@inproceedings{koch2015siamese,
  title={{Siamese Neural Networks for One-Shot Image Recognition}},
  author={Koch, Gregory and Zemel, Richard and Salakhutdinov, Ruslan},
  booktitle={ICML Deep Learning Workshop},
  address={Lille, France},
  month=jul,
  volume={2},
  year={2015}
}

@book{goodfellow2016deep,
  title={{Deep Learning}},
  author={Goodfellow, Ian and Bengio, Yoshua and Courville, Aaron and Bengio, Yoshua},
  volume={1},
  year={2016},
  publisher={MIT press Cambridge}
}

@inproceedings{cvrcek2006study,
  title={{A Study on the Value of Location Privacy}},
  author={Cvrcek, Dan and Kumpost, Marek and Matyas, Vashek and Danezis, George},
  booktitle={ACM WPES},
  pages={109--118},
  address={Alexandria, VA, USA},
  month=oct,
  year={2006}
}

@inproceedings{brush2010exploring,
  title={{Exploring End-User Preferences for Location Obfuscation, Location-Based Services, and the Value of Location}},
  author={Brush, AJ and Krumm, John and Scott, James},
  booktitle={ACM UbiComp},
  pages={95--104},
  year={2010},
  month=sep,
  address={Copenhagen, Denmark}
}

@inproceedings{shih2015privacy,
  title={{Privacy Tipping Points in Smartphones Privacy Preferences}},
  author={Shih, Fuming and Liccardi, Ilaria and Weitzner, Daniel},
  booktitle={ACM CHI},
  pages={807--816},
  month=apr,
  year={2015},
  address={Seoul, Republic of Korea}
}

@article{song2010limits, 
  title={{Limits of Predictability in Human Mobility}},
  author={Song, Chaoming and Qu, Zehui and Blumm, Nicholas and Barab{\'a}si, Albert-L{\'a}szl{\'o}}, 
  journal={Science}, 
  volume={327}, 
  number={5968}, 
  pages={1018--1021}, 
  year={2010}, 
  publisher={American Association for the Advancement of Science}
}

@ARTICLE{kailath1967divergence,
author={Thomas Kailath},
journal={IEEE Transactions on Communication Technology},
title={{The Divergence and Bhattacharyya Distance Measures in Signal Selection}},
year={1967},
volume={15},
number={1},
pages={52-60},
}

@INPROCEEDINGS{perez2019magID,
 author   = {Beatrice Perez and Marco Musolesi and Gianluca Stringhini},
 title    = {{Fatal Attraction: Identifying Mobile Devices Through Electromagnetic Emissions}},
 booktitle   = {WiSec},
 year     = {2019}
}

@INPROCEEDINGS{dey2014accelprint,
  author = {Dey, Sanorita and Roy, Nirupam and Xu, Wenyuan and Choudhury, Romit
	Roy and Nelakuditi, Srihari},
  title = {{AccelPrint: Imperfections of Accelerometers Make Smartphones Trackable}},
  year = {2014},
  booktitle = {NDSS '14},
  address = {San Diego, CA, USA}
}

@MISC{sensor-overview,
  howpublished = {\url{https://developer.android.com/about/dashboards/index.html}},
  note = {Accessed: 2024-07-01},
  title = {Sensors Overview}}

@book{brook1988signal,
publisher = {Edward Arnold},
isbn = {0713135646},
year = {1988},
title = {Signal processing: principles and applications},
address = {London, England, UK},
author = {Brook, D. and Wynne, R.J.},
}

@article{rabiner1978digital,
  title={{Digital Processing of Speech Signal}},
  author={Rabiner, Lawrence R},
  journal={Digital Processing of Speech Signal},
  year={1978},
  publisher={Prentice-hall}
}

@article{internetcomputing,
	author={Andrew T. Campbell and Shane B. Eisenman and Nicholas D. Lane and Emiliano Miluzzo and Ronald Peterson and Hong Lu and Xiao Zheng and Mirco Musolesi and Kristof Fodor and Gahng-Seop Ahn},
	title={{The Rise of People-Centric Sensing}},
	journal={IEEE Internet Computing Special Issue on Mesh Networks},
	month={June/July},
	year={2008}
}

@INPROCEEDINGS{zhang2019sensorid,
  author={Zhang, Jiexin and Beresford, Alastair R. and Sheret, Ian},
  booktitle={IEEE S\&P}, 
  title={SensorID: Sensor Calibration Fingerprinting for Smartphones}, 
  year={2019},
  volume={},
  number={},
  pages={638-655}}

@article{kumar2023device,
  title={Device fingerprinting for cyber-physical systems: a survey},
  author={Kumar, Vijay and Paul, Kolin},
  journal={ACM Computing Surveys},
  volume={55},
  number={14s},
  pages={1--41},
  year={2023},
  publisher={ACM New York, NY}
}

@article{mohamed2023istelan,
  title={Istelan: Disclosing sensitive user information by mobile magnetometer from finger touches},
  author={Mohamed, Reham and Farrukh, Habiba and Lu, Yidong and Wang, He and Celik, Z Berkay},
  journal={Proceedings on Privacy Enhancing Technologies},
  year={2023}
}

@article{yapar2023real,
  title={Real-time outdoor localization using radio maps: A deep learning approach},
  author={Yapar, {\c{C}}a{\u{g}}kan and Levie, Ron and Kutyniok, Gitta and Caire, Giuseppe},
  journal={IEEE Transactions on Wireless Communications},
  year={2023},
  volume = {22},
issue = {12},
  publisher={IEEE}
}

@inproceedings{noulas2009inferring,
  title={Inferring interests from mobility and social interactions},
  author={Noulas, Anastasios and Musolesi, Mirco and Pontil, Massimiliano and Mascolo, Cecilia},
  booktitle={NIPS Workshop on Analyzing Networks and Learning with Graphs},
  pages={2--88},
  year={2009}
}

\end{document}